\documentclass[aps,prd,preprint,tightenlines,groupedaddress,nofootinbib,showpacs,byrevtex]{revtex4}
\usepackage{amssymb,latexsym}
\usepackage{amsmath,amsbsy}
\usepackage{epsfig,bm}
\usepackage{graphicx,comment}
\DeclareMathOperator{\tr}{tr}

\begin{document}
\def\a{{\alpha}}
\def\b{{\beta}}
\def\d{{\delta}}
\def\D{{\Delta}}
\def\e{{\varepsilon}}
\def\g{{\gamma}}
\def\G{{\Gamma}}
\def\k{{\kappa}}
\def\l{{\lambda}}
\def\L{{\Lambda}}
\def\m{{\mu}}
\def\n{{\nu}}
\def\o{{\omega}}
\def\O{{\Omega}}
\def\S{{\Sigma}}
\def\s{{\sigma}}
\def\th{{\theta}}

\def\ol#1{{\overline{#1}}}

\def\Dslash{D\hskip-0.65em /}

\def\CPT{{$\chi$PT}}
\def\QCPT{{Q$\chi$PT}}
\def\PQCPT{{PQ$\chi$PT}}
\def\tr{\text{tr}}
\def\str{\text{str}}
\def\diag{\text{diag}}
\def\order{{\mathcal O}}

\def\cC{{\mathcal C}}
\def\cB{{\mathcal B}}
\def\cT{{\mathcal T}}
\def\cQ{{\mathcal Q}}
\def\cL{{\mathcal L}}
\def\cO{{\mathcal O}}
\def\cA{{\mathcal A}}
\def\cQ{{\mathcal Q}}
\def\cR{{\mathcal R}}
\def\cH{{\mathcal H}}
\def\cW{{\mathcal W}}
\def\cM{{\mathcal M}}
\def\cJ{{\mathcal J}}
\def\cF{{\mathcal F}}
\def\cK{{\mathcal K}}
\def\cY{{\mathcal Y}}

\def\eqref#1{{(\ref{#1})}}

 
\title{Baryons with Ginsparg-Wilson quarks in a staggered sea}
\author{ Brian C.~Tiburzi}
\email[]{bctiburz@phy.duke.edu}
\affiliation{Department of Physics\\
Duke University\\
P.O.~Box 90305\\
Durham, NC 27708-0305}

\date{\today}


\begin{abstract}
We determine the masses and magnetic moments of the octet baryons
in chiral perturbation theory formulated for a mixed lattice action 
of Ginsparg-Wilson valence quarks and staggered sea quarks. 
Taste-symmetry breaking does not occur at next-to-leading order in the combined lattice-spacing and chiral expansion. 
Expressions derived for masses and magnetic moments are required for addressing lattice artifacts in 
mixed-action simulations of these observables.  
\end{abstract}

\maketitle

\section{Introduction}

Lattice QCD has made impressive progress. As fully dynamical
simulations for a wide range of observables are carried out at lighter quark masses, 
our confidence in lattice QCD as a predictive tool grows. There are a number of 
fermion discretizations used on the lattice. These are at various stages of development, 
and are fraught with disparate difficulties.  Further confidence in lattice QCD 
will eventually be built through comparisons of simulations employing different lattice fermions.

Presently lattice QCD studies using dynamical staggered fermions~\cite{Susskind:1976jm} reach 
smaller quark masses as compared with other lattice fermions. The publicly available MILC configurations~\cite{Bernard:2001av}, 
moreover, have launched dynamical staggered fermions as readily accessible 
for lattice calculations~\cite{Davies:2003ik,Aubin:2004wf,Aubin:2004ck,Aubin:2004fs,Aubin:2004ej,Aubin:2005ar}. 
The fourth-root trick is employed by these simulations to reduce the number of so-called taste degrees of freedom. 
In the continuum limit, there are no taste-changing interactions and the trick is kosher. On the lattice 
at finite $a$, however, the question of locality emerges and the trick remains controversial, 
for various  recent investigations 
see~\cite{Durr:2003xs,Bunk:2004br,Follana:2004sz,Durr:2004as,Durr:2004ta,Maresca:2004me,Adams:2004mf,Wong:2004nk,Shamir:2004zc}.
Nonetheless, even with improved staggered actions the discretization effects are surprisingly large. 
Chiral perturbation theory has been extended to staggered actions to control the systematic errors 
associated with the continuum extrapolation~\cite{Lee:1999zx,Aubin:2003mg,Aubin:2003uc,Sharpe:2004is,Aubin:2004xd,VandeWater:2005uq}.

Simulations that employ Ginsparg-Wilson fermions~\cite{Ginsparg:1981bj}, on the other hand, do not suffer
from the precarious theoretical situation of staggered fermions.\footnote{%
In this work, we assume the validity of the fourth-root trick. 
Comparing lattice data with the behavior of observables predicted from staggered chiral perturbation theory
could yield empirical evidence for or against the trick. 
}
These fermions, moreover, have
an exact chiral symmetry (which in practice is limited by how well the overlap 
operator~\cite{Narayanan:1992wx,Narayanan:1993sk,Narayanan:1994gw} is approximated, 
or how well the domain wall fermion~\cite{Kaplan:1992bt,Shamir:1993zy,Furman:1994ky} is realized). 
Computationally, however, Ginsparg-Wilson fermions are numerically quite costly. Mid-ground between these two 
lattice fermions can be found. Ginsparg-Wilson quarks can be utilized for the valence quarks, where 
they are computationally less demanding.  One can then calculate correlation functions in the background of  
the existing staggered sea of the MILC configurations. 
This efficacious solution has costs comparable to quenched Ginsparg-Wilson simulations, and
numerical investigations in mixed-action lattice QCD  have been 
undertaken~\cite{Renner:2004ck,Bowler:2004hs,Bonnet:2004fr,Beane:2005rj}. 
Mixed-action chiral perturbation theory has recently been formulated in the meson sector~\cite{Bar:2005tu}
to analyze the lattice-spacing dependence of meson observables.

In this work, we construct the baryon chiral Lagrangian
for a mixed lattice action consisting of Ginsparg-Wilson
valence quarks and staggered sea quarks. We apply this 
Lagrangian to the computation of octet baryon masses and
magnetic moments. 
To address the effects of finite lattice spacing, 
one formulates the underlying lattice theory and matches it onto a chiral effective theory~\cite{Sharpe:1998xm}. 
To do so, we utilize a dual expansion in the quark masses and lattice spacing. 
We assume a hierarchy of energy scales
\begin{equation}
m_q \ll \L_{QCD} \ll \frac{1}{a}
,\end{equation}
and further choose the power counting scheme 
\begin{equation} \label{eqn:pc}
\e^2 \sim 
\begin{cases}
 m_q / \L_{QCD} \\
 a^2   \L_{QCD}^2
\end{cases}
,\end{equation}
which is currently relevant for simulations employing 
improved staggered quarks~\cite{Aubin:2004fs}.
The resulting quark mass and lattice 
spacing dependent expressions will be useful for the
analysis of numerical results from simulations of baryon observables in mixed-action lattice QCD.

Our presentation has the following organization. 
First we briefly review the Symanzik Lagrangian for 
the mixed lattice action in Section~\ref{s:In}.
Here we recall the form of the chiral Lagrangian in the meson sector~\cite{Bar:2005tu},
and cite the relevant details for our calculation. 
The baryons are then included in the theory. 
In Section~\ref{s:magmom},
we determine the magnetic moments to $\cO(\e)$ in the combined expansion; 
while in Section~\ref{s:mass}, we calculate the masses of the octet baryons
up to $\cO(\e^3)$ . These calculations at their respective orders
include the leading non-analytic dependence on the quark masses, and are shown
to be devoid of taste-symmetry breaking.  
Taste-symmetry breaking interactions contribute at $\cO(\e^4)$
from loops that scale generically as  $\cO(a^2 m_q \log m_q  )$ and $\cO(m_q^2 \log m_q)$. 
A summary ends the paper (Section~\ref{s:summy}). 
For completeness we include a discussion of the finite volume corrections 
in the Appendix.

\section{Chiral Lagrangian} \label{s:In}

Before including baryons into the mixed-action chiral Lagrangian for Ginsparg-Wilson
valence quarks and staggered sea quarks, we 
first give a brief review of the Symanzik Lagrangian. 
Next we recall the form of the chiral effective theory in the meson sector, 
and list results pertinent for the calculations in this paper.  The relevant
pieces of the baryon Lagrangian are then detailed.

\subsection{Symanzik Lagrangian} 

The lattice action can be described in terms of a continuum effective 
field theory. This effective theory is described by the Symanzik action~\cite{Symanzik:1983dc,Symanzik:1983gh}, 
which is built from continuum operators and is based on the symmetries of the underlying lattice theory. 
The Symanzik Lagrangian is organized in powers of the lattice spacing $a$, namely
\begin{equation}
\cL_{\text{Sym}} = 
\cL 
+ 
a \, \cL^{(5)} 
+ 
a^2 \, \cL^{(6)} 
+ 
a^3 \, \cL^{(7)}
+
\dots \label{eq:syman}
,\end{equation}
where $\cL^{(n)}$ represents the contribution from dimension-$n$ operators.\footnote{%
One should note that not all $a$-dependence is parametrized in Eq.~\eqref{eq:syman}. 
The coefficients of operators in $\cL^{(n)}$ depend on the gauge coupling, and hence can have a 
weak logarithmic dependence on $a$. 
}  
The symmetries of the mixed lattice action 
are respected by the Symanzik Lagrangian $\cL_{\text{Sym}}$ order-by-order in $a$. 
In the continuum limit, $a \to 0$, only the operators of $\cL$ survive.
We consider here the case of a mixed action in partially quenched QCD (PQQCD). 
This type of action allows not only for the valence and sea quarks to have different masses, 
but to be different types of lattice fermions.  An important feature of mixed action theories 
concerns the general lack of symmetry between the valence and sea sectors~\cite{Bar:2002nr}. 
Because different types of lattice fermions are used in each sector of the theory, the flavor symmetry
of the mixed action is generally a direct product of the separate symmetries in the valence and sea sectors.

The underlying lattice action we consider is built from three flavors of Ginsparg-Wilson valence quarks and 
three flavors of staggered sea quarks. In the continuum limit, the Lagrangian $\cL$ is just the 
partially quenched action,\footnote{%
We use the super-symmetric formulation of partially quenched theories that stems back to~\cite{Morel:1987xk}.
One could equivalently use the replica method~\cite{Damgaard:2000gh}.
}
 namely
\begin{equation}
\cL = \ol Q \Dslash \, Q + \ol Q m_q Q,
\end{equation} 
where the quark fields appear in the vector
\begin{equation}
Q = ( u, d, s, j_1, j_2, j_3, j_4, l_1, l_2, l_3, l_4, r_1, r_2, r_3, r_4, \tilde{u}, \tilde{d}, \tilde{s} )^T
.\end{equation}
Notice that fermion doubling has produced four tastes for each flavor $(j,l,r)$
of staggered quark.
In a partially quenched generalization of the isospin limit, the mass matrix is given by
\begin{equation}
m_q = \diag (m_u, m_u, m_s, m_j \xi_I, m_j \xi_I , m_r \xi_I , m_u, m_u, m_s ) 
,\end{equation}
with $\xi_I$ as the $4$ x $4$ taste identity matrix. 
In the massless limit, the Lagrangian $\cL$ has a graded chiral symmetry of the form 
$SU(15|3)_L \otimes SU(15|3)_R$. 
For our discussion below, it is useful to define projection operators for the 
valence ($V$) and sea ($S$) sectors of the theory:
$\mathcal{P}_V = \diag ( 1, 1, 1, 0, \ldots, 0, 1, 1, 1)$, 
and 
$\mathcal{P}_S = (0, 0, 0, 1 \xi_I, 1 \xi_I, 1 \xi_I, 0, 0, 0)$.

The Ginsparg-Wilson sector of the theory possesses an exact chiral symmetry in the limit
of zero quark mass~\cite{Luscher:1998pq}.  Thus there can be no operators in $\cL^{(5)}$
involving only valence quarks because the only dimension-$5$ operator (after field redefinitions~\cite{Luscher:1996sc})
is a chiral symmetry breaking quark bilinear. Also there are no dimension-$5$
operators built from just staggered quark fields~\cite{Sharpe:1993ng,Luo:1996vt}. 
Finally the symmetries of the mixed action forbid bilinear operators formed from one valence and one sea quark. 
Thus $\cL^{(5)} = 0$.

Next we consider the dimension-$6$ operators, of which there are purely gluonic operators, quark bilinears and 
four-quark operators. The gluonic operators can be omitted from consideration. This is because they transform 
as singlets under chiral transformations in the valence and sea sector, and their contribution in the effective 
theory will be identical to those from two-quark and four-quark operators of dimension-$6$ 
that do not break the $SU(15|3)_L \otimes SU(15|3)_R$ chiral symmetry. 
We decompose the dimension-$6$ operators into three classes as follows
\begin{equation}
\cL^{(6)} = \cL^{(6)}_{\text{val}} + \cL^{(6)}_{\text{sea}} + \cL^{(6)}_{\text{mix}}
.\end{equation}
The first term $\cL^{(6)}_{\text{val}}$ consists of operators formed from only valence quark fields. 
When the valence quark masses are zero, these operators have an $SU(3|3)_L \otimes SU(3|3)_R$ 
chiral symmetry which is reduced to a
vector symmetry when the valence masses are turned on.  
There is also a term in $\cL^{(6)}_{\text{val}}$ which breaks the $SO(4)$ rotational invariance
of Euclidean space down to the hypercubic group $SW_4$. The dimension-$6$ terms in $\cL^{(6)}_{\text{val}}$  
for unquenched theories have been detailed long ago~\cite{Sheikholeslami:1985ij}, 
for a recent discussion that addresses Ginsparg-Wilson valence quarks specifically, see~\cite{Bar:2003mh}.

The second term $\cL^{(6)}_{\text{sea}}$ consists of operators formed from only
sea quark fields. The case of one staggered flavor (corresponding to four tastes) was considered in~\cite{Luo:1997tt,Lee:1999zx}. 
More recently these results were generalized to the case of multiple flavors in~\cite{Aubin:2003mg,Sharpe:2004is}. 
Terms that do not break taste-symmetry have the same form as operators in $\cL^{(6)}_{\text{val}}$. These operators 
have an $SU(12)_L \otimes SU(12)_R$ 
chiral symmetry in the massless sea quark limit, and a vector symmetry away from zero mass. 
There is a single term which reduces the rotational symmetry of taste-symmetric terms down to $SW_4$. 
When one considers taste-symmetry breaking terms, one has four-quark operators of the form 
\begin{equation}
\Big[ \ol Q ( \Gamma_A \otimes \Gamma^*_B) \mathcal{P}_S  Q \Big]
\Big[ \ol Q (\Gamma_A \otimes \Gamma^*_B ) \mathcal{P}_S Q \Big]
\end{equation}
where $\Gamma_A$ denotes a Dirac matrix that acts on the spin indices of the quark fields, 
and $\Gamma^{*}_B$ denotes a generator of $SU(4)$ taste that acts on the taste indices of the quark fields.
We treat the sum over staggered flavors as implicit. The presence of explicit $\Gamma^*_B$ matrices breaks 
the taste-symmetry. When the collective labels $A$ and $B$ are unrelated, the operator respects $SO(4)$ 
rotational invariance. 
On the other hand, when the Lorentz indices contained in label $A$ are contracted with those in $B$, both 
taste and rotational symmetries are broken and consequently the distinction between the two is blurred.

Finally operators in $\cL^{(6)}_{\text{mix}}$ are operators that involve both valence and sea quark fields. 
Because of the mixed action symmetry, all of these terms are four-quark operators consisting of a product of 
a bilinear of valence quarks  with a bilinear of sea quarks.  These operators were constructed in~\cite{Bar:2005tu},
where it was shown that only taste-singlet sea-quark bilinears are present. The possible Dirac structures of these bilinears 
are constrained by the exact chiral symmetry required in the valence sector, and the axial symmetry
required in the sea sector.  Only vector and axial-vector bilinears are consistent with both symmetries.
Ignoring color structure which has no bearing in the construction of the chiral 
Lagrangian, we have two terms in $\cL^{(6)}_{\text{mix}}$
\begin{equation}
\cL^{(6)}_{\text{mix}} 
= 
C_{\text{mix}}^V
\left( \ol Q \gamma_\mu \mathcal{P}_V Q \right) \left( \ol Q \gamma_\mu \mathcal{P}_S  Q \right)
+
C_{\text{mix}}^A
\left( \ol Q \gamma_\mu \gamma_5 \mathcal{P}_V Q \right) \left( \ol Q \gamma_\mu \gamma_5 \mathcal{P}_S Q \right)
.\end{equation}
These terms have a $SU(3|3)_L \otimes SU(3|3)_R \otimes SU(12)_L \otimes SU(12)_R$ chiral symmetry.

Finally we consider the terms in $\cL^{(7)}$. We do not consider operators that have a quark mass insertion
because these will necessarily be higher order in our power counting. There are no operators that consist only 
of Ginsparg-Wilson valence quarks because we cannot build parity even, chirally symmetric operators of 
dimension-$7$. As discussed in~\cite{Sharpe:2004is}, there are no dimension-$7$ operators built solely 
from staggered quark fields due to the staggered axial symmetry. Finally any operators in $\cL^{(7)}$ 
that consist of both staggered quarks and Ginsparg-Wilson quarks must occur as a product of bilinears 
of each fermion type. Again this is because there is no symmetry relating the two sectors of the mixed-action theory. 
There are no such bilinears, however, because one cannot write down a dimension-$7$ operator consisting of
a chirally symmetric Ginsparg-Wilson bilinear and an axially symmetric staggered bilinear. Thus $\cL^{(7)} = 0$.

\subsection{Mesons}

In this section, we review the construction of chiral perturbation theory in the meson sector of this partially-quenched 
mixed-action theory.
Partially quenched chiral perturbation theory was developed through a series of 
papers~\cite{Bernard:1994sv,Sharpe:1997by,Golterman:1998st,Sharpe:2000bc,Sharpe:2001fh}. 
The construction of the chiral effective theory for Ginsparg-Wilson valence quarks and staggered sea quarks 
was carried out in~\cite{Bar:2005tu}. We will focus only on the ingredients of this theory necessary for 
determining baryon properties at next-to-leading order.

As commented above, in the massless and continuum limit the Symanzik action has a graded chiral symmetry $SU(15|3)_L \otimes SU(15|3)_R$.
We expect this symmetry to be spontaneously broken down to $SU(15|3)_V$ in analogy with QCD.  Thus we can build an effective 
theory of mixed action PQQCD written in terms of the pseudo-Goldstone modes that emerge from spontaneous chiral 
symmetry breaking. 
These modes acquire masses from the explicit chiral symmetry breaking introduced by the quark mass term in the continuum 
action. Additionally we shall see below that the $a$-dependent terms in the 
Symanzik action also generate masses for some of these modes. 
We shall generically call the pseudo-Goldstone modes mesons, and collect them in an $U(15|3)$ matrix exponential $\Sigma$
\begin{equation}
\Sigma = e^{2 i \Phi / f} \equiv \xi^2
,\end{equation}
that is written in terms of the meson matrix $\Phi$. 
We work almost exclusively in the quark basis due to the fact that mesons 
composed of two sea quarks will enter our next-to-leading order calculations 
only as taste singlets.

At $\cO(\e^2)$ in the combined lattice spacing and chiral expansion the effective Lagrangian of partially quenched
chiral perturbation theory (\PQCPT) that describes the mesons has the form~\cite{Bar:2005tu}
\begin{equation} \label{eq:Llead}
\cL =  
\frac{f^2}{8}
\str \left( \partial_\mu\Sigma^\dagger \partial_\mu\Sigma\right)
    - \l \,\str\left(m_q\Sigma^\dagger+m_q^\dagger\Sigma\right)
    + \frac{1}{6} \mu_0^2 \, (\str \, \Phi)^2
    + a^2 \mathcal{V}    
.\end{equation}
With these conventions, the pion-decay constant $f = 132 \, \texttt{MeV}$. 
The potential $\mathcal{V}$ contains the effects of dimension-$6$ operators
in the Symanzik action.  It can be decomposed into three terms~\cite{Bar:2005tu}
\begin{equation}
\mathcal{V} = \mathcal{U}_S + \mathcal{U}'_S + \mathcal{U}_{V}
.\end{equation}
Above $\mathcal{U}_S$ and $\mathcal{U}'_S$ are taste-symmetry breaking potentials
involving single trace and two trace operators, respectively~\cite{Lee:1999zx,Aubin:2003mg}. 
These potentials involve only the mesons composed of two sea quarks and arise from 
$\cL^{(6)}_{\text{sea}}$ in the Symanzik action. The potential $\mathcal{U}_V$
contains all terms that stem from Symanzik operators involving the valence quark fields.
These operators are in $\cL^{(6)}_{\text{val}}$ and $\cL^{(6)}_{\text{mix}}$.  
As shown in~\cite{Bar:2005tu}, there is only one term in the valence potential
\begin{equation}
\mathcal{U}_V = - a^2 \, C_{\text{mix}} \, \str \left( \tau_3 \Sigma \tau_3 \Sigma^\dagger \right)
,\end{equation}
where $\tau_3 = \mathcal{P}_S - \mathcal{P}_V$. This term only acts on mesons composed of one valence 
quark and one sea quark. At this order there are no contributions from operators that act on valence-valence mesons
as their effects lead to $a^2$ renormalization of the lowest order parameters~\cite{Bar:2003mh}.

Working to tree-level, we can determine the meson masses needed below for the calculation of 
baryon observables.
When one expands the Lagrangian in Eq.~\eqref{eq:Llead} to leading order, one finds that mesons
formed of two Ginsparg-Wilson quarks $Q \ol Q {}'$ have mass
\begin{equation} \label{eq:GWGW}
m_{QQ'}^2 = \frac{4 \lambda }{f^2} ( m_Q + m_{Q'})  
,\end{equation}
which accordingly vanishes in the chiral limit. 
Mesons consisting of a staggered quark $Q_{i}$, i.e. of flavor $Q$ and quark taste $i$, 
and a Ginsparg-Wilson quark $\ol Q {}'$ have masses
\begin{equation} \label{eq:SGW}
m_{Q_i Q'}^2 = \frac{4 \lambda }{f^2} ( m_Q + m_{Q'}) + \frac{16 \, a^2 \, C_{\text{mix}}}{f^2} 
.\end{equation}
Notice that these masses do not depend on the quark taste. 
Masses of mixed mesons do not vanish in the chiral limit because the full chiral symmetry of 
$\cL$ is explicitly broken by $\cL^{(6)}_{\text{mix}}$ down to the mixed action chiral symmetry. 
The final mesons relevant 
to baryon observables at next-to-leading order are those with two staggered quarks 
in a flavor-neutral, taste-singlet combination. In the quark basis, their masses are given by
\begin{equation} \label{eq:SS}
m_{QQ}^2 = \frac{8 \lambda }{f^2} m_Q  + \frac{64 \, a^2}{f^2} (C_3 + C_4)
,\end{equation}
where $C_3$ and $C_4$ are parameters entering the potential $\mathcal{U}_S$.

Notice that in Eq.~\eqref{eq:Llead} the matrix $\Phi$ is not supertraceless, and we have included the singlet
mass parameter $\mu_0$. This is a device to derive the form of the flavor-neutral propagators~\cite{Sharpe:2001fh}. 
Partially quenched theories have a $U(1)_A$ anomaly which renders the singlet heavy. Thus the form of 
flavor-neutral propagators can be derived in the limit that $\mu_0 \to \infty$, and the singlet [here of $SU(15|3)_V$] 
is integrated out of the low-energy theory. For the staggered flavor-neutral propagators, the analysis
has been carried out in~\cite{Aubin:2003mg}, including the inclusion of $1/4$-factors corresponding to the fourth-root trick. 
There it was shown that different flavor-neutral propagators exist for the taste singlet, vector and axial-vector 
channels. To the order we work, the only flavor neutrals required will be composed of two Ginsparg-Wilson valence quarks. 
Thus as with the meson mass and decay constant in mixed action \PQCPT~\cite{Bar:2005tu}, 
only the taste-singlet flavor-neutral propagator
is required. For $a,b = u, d, s$, this propagator is given by
\begin{equation}
 \label{eq:singlet}
{\cal G}_{\eta_a \eta_b} =
        \frac{\delta_{ab}}{k^2 + m^2_{aa}}
        - \frac{1}{3} \frac{ \left(k^2 + m^2_{jj}
            \right) \left( k^2 + m^2_{rr} \right)}
            {\left(k^2 + m^2_{aa} \right)
             \left(k^2 + m^2_{bb} \right)
             \left(k^2 + m^2_X \right)}\, ,
\end{equation}
where the masses of the valence-valence mesons, $m_{aa}^2$ and $m_{bb}^2$, are given in Eq.~\eqref{eq:GWGW}, 
while those of the sea-sea mesons, $m_{jj}^2$ and $m_{rr}^2$, are given in Eq.~\eqref{eq:SS}. 
The mass $m_X$ is defined as $m_X^2 =\frac{1}{3}\left(m^2_{jj} + 2 m^2_{rr}\right)$.  The flavor-neutral  
propagator can be conveniently rewritten as
\begin{equation}
{\cal G}_{\eta_a \eta_b} =
         \d_{ab} P_a +
         {\cal H}_{ab}\left(P_a,P_b,P_X\right),
\end{equation}
where
\begin{eqnarray}
     P_a &=& \frac{1}{k^2 + m^2_{aa}},\ 
     P_b = \frac{1}{k^2 + m^2_{bb}},\ 
     P_X = \frac{1}{k^2 + m^2_X}, \,
\nonumber\\
\nonumber\\
\nonumber\\
     {\cal H}_{ab}\left(A,B,C\right) &=& 
           -\frac{1}{3}\left[
             \frac{\left( m^2_{jj} - m^2_{aa}\right)
                   \left( m^2_{rr} - m^2_{aa}\right)}
                  {\left( m^2_{aa} - m^2_{bb}\right)
                   \left( m^2_{aa} - m^2_X\right)}
                 A
            +\frac{\left( m^2_{jj} - m^2_{bb}\right)
                   \left( m^2_{rr} - m^2_{bb}\right)}
                  {\left( m^2_{bb} - m^2_{aa}\right)
                   \left( m^2_{bb} - m^2_{X}\right)}
                 B \right.\, 
\nonumber\\
&&\qquad\quad\left.
            +\frac{\left( m^2_{jj} - m^2_{X}\right)
                   \left( m^2_{rr} - m^2_{X}\right)}
                  {\left( m^2_X - m^2_{aa}\right)
                   \left( m^2_X - m^2_{bb}\right)}
                 C\ \right].
\label{eq:Hfunction}
\end{eqnarray}

In the limit that $b \to a$, we require a separate form of the flavor-neutral propagator
to handle the double pole. The functional form, however, can be related to a derivative of the 
single pole form, namely
\begin{equation}
\cH_{aa}(A, A, C) 
= 
-\frac{1}{3} 
\left[ 
\frac{\partial}{\partial m_{aa}^2} \frac{(m_{jj}^2 - m_{aa}^2)(m_{rr}^2 - m_{aa}^2)}{(m_{aa}^2 - m_X^2)} A
+
\frac{(m_{jj}^2 - m_X^2)(m_{rr}^2 - m_X^2)}{(m_X^2 - m_{aa}^2)^2} C
\right]
,\end{equation} 
keeping in mind that $A = A(m_{aa}^2)$.

\subsection{Baryons}

Having reviewed the mixed action Symanzik Lagrangian and the relevant pieces of meson \PQCPT\ at finite $a$, 
we now extend mixed action \PQCPT\ to the baryon sector. In the continuum limit, 
the flavor symmetry group of the mixed action theory is $SU(15|3)_V$. There are no taste-symmetry 
breaking interactions present when the lattice spacing is zero. 
Thus we can construct spin-$1/2$ and spin-$3/2$ baryon flavor
tensors in the usual fashion~\cite{Savage:2001dy,Chen:2001yi,Beane:2002vq}.\footnote{%
This is a useful way to proceed because the external states consist of three Ginsparg-Wilson quarks
and have a one-to-one correspondence with the baryon interpolating fields used on the lattice. 
On the other hand, baryons formed from quark tastes are complicatedly related to the
usual staggered quark interpolating operators used on the lattice.
}
In general we can find the dimensionality of these multiplets in $SU(M|N)_V$
using Young super-tableaux~\cite{BahaBalantekin:1980qy,BahaBalantekin:1980pp}.
For the spin-$1/2$ multiplet $\cB^{ijk}$, we have a 
\begin{equation}
\frac{1}{3} \Big[(M+1) M (M-1) + (N+1) N (N-1)\Big] + M N ( M + N )
\end{equation}
dimensional representation of $SU(M|N)_V$. While the spin-$3/2$ multiplet $\cT^{ijk}_\mu$
furnishes a
\begin{equation}
\frac{1}{4} M ( M + 1) + \frac{1}{12} M ( M + 1) ( 2 M + 1) + \frac{1}{2} M N ( M + N ) 
+ \frac{1}{2} (N - 1) (N-2) - \d_{N,0} 
\end{equation}
dimensional representation.

For the case at hand, $SU(15|3)_V$ \PQCPT, we embed the spin-$\frac{1}{2}$ baryons 
in the $\bf{1938}$-dimensional super-multiplet $\cB^{ijk}$. 
The spin-$\frac{3}{2}$ baryons are embedded in the $\bf{1086}$-dimensional super-multiplet 
$\cT_\mu^{ijk}$. For baryon \PQCPT\ at next-to-leading order (and indeed at next-to-next-to-leading
order as well), we shall need the states in these multiplets consisting of at most one sea quark
or at most one ghost quark. To this end, we decompose the irreducible representations
of $SU(15|3)_V$ into irreducible representations of 
$SU(3)_{\text{val}} \otimes SU(12)_{\text{sea}} \otimes SU(3)_{\text{ghost}}$~\cite{Hurni:1981ki}. 
To describe these super-algebra multiplets we refer to their floors, where the floor
number coincides with the number of bosonic ghost quarks contained in states. The ground floor is synonymous
with zero ghost quarks. Additionally we refer to the levels of the multiplet~\cite{Chen:2001yi} to distinguish
between baryon states with differing numbers of sea quarks. Level $A$ baryons do not have a sea quark, level $B$
baryons have one sea quark,  and so on.

The ground floor, level $A$ of the multiplet $\cB^{ijk}$ consists of baryons that 
transform as a $(\bm{8}, \bm{1}, \bm{1})$ under
$SU(3)_{\text{val}} \otimes SU(12)_{\text{sea}} \otimes SU(3)_{\text{ghost}}$. 
These are the octet baryons, and are embedded in the tensor $\cB^{ijk}$ in the standard 
way when all of the indices are restricted to $1$--$3$~\cite{Labrenz:1996jy}.
The first floor, level $A$ of $\cB^{ijk}$ transforms as a 
$(\bm{6}, \bm{1}, \bm{3}) \oplus (\bm{\ol 3}, \bm{1}, \bm{3})$.
These states have been constructed explicitly in~\cite{Chen:2001yi}.
The ground floor, level $B$ of the $\cB^{ijk}$ multiplet transforms 
as a $(\bm{6}, \bm{12}, \bm{1}) \oplus (\bm{\ol 3}, \bm{12}, \bm{1})$. 
The states constructed in~\cite{Chen:2001yi} can be used with minimal modifications. 
One merely must re-index so that the range of $q_{\text{sea}}$ is extended from $1$--$3$ to $1$--$12$,
or merely attach a taste index to each flavor tensor.

The situation is the same with respect to the spin-$3/2$ baryon tensor $\cT_\mu^{ijk}$. Either
the states have already been constructed in~\cite{Chen:2001yi}, or it is trivial to extend to the present case 
those which have not.
The ground floor, level $A$ of the spin-$3/2$ multiplet transforms as 
a $(\bm{10}, \bm{1}, \bm{1})$ under $SU(3)_{\text{val}} \otimes SU(12)_{\text{sea}} \otimes SU(3)_{\text{ghost}}$,
and consists of the decuplet baryons which are embedded in the usual fashion. 
The first floor, level $A$ of $\cT_\mu^{ijk}$ transforms as
a $(\bm{6}, \bm{1}, \bm{3})$, while the ground floor, level $B$ transforms as a  
$(\bm{6}, \bm{12}, \bm{1})$.
Thus the total number of baryon states relevant for calculating loop diagrams up to next-to-next-to-leading order
is $243$, which is considerably smaller than the total number of baryons in the theory $3024$.

To $\cO(\e^2)$, the free Lagrangian for the $\cB^{ijk}$ and $\cT^{ijk}_\mu$ fields 
retains the same form as in quenched and partially quenched theories~\cite{Labrenz:1996jy,Chen:2001yi}
with the addition of new lattice-spacing dependent terms,
\begin{eqnarray} \label{eqn:L}
  {\mathcal L}
  &=&
   i\left(\ol\cB v\cdot{\mathcal D}\cB\right)
  -2\a_M\left(\ol\cB \cB{\mathcal M}_+\right)
  -2\b_M\left(\ol\cB {\mathcal M}_+\cB\right)
  -\frac{1}{2}\sigma_M\left(\ol\cB\cB\right)\str\left({\mathcal M}_+\right)
  + a^2 \mathcal{V}_{\cB}
                              \nonumber \\
   &&+i\left(\ol\cT_\mu v\cdot{\mathcal D}\cT_\mu\right)
    +\D\left(\ol\cT_\mu\cT_\mu\right)
    +2\g_M\left(\ol\cT_\mu {\mathcal M}_+\cT_\mu\right)
    -\frac{1}{2}\ol\sigma_M\left(\ol\cT_\mu\cT_\mu\right)\str\left({\mathcal M}_+\right) 
   + a^2 \mathcal{V}_{\cT}. \notag \\
\end{eqnarray}
The baryon potentials $\mathcal{V}_\cB$ and $\mathcal{V}_\cT$
arise from the operators in $\cL^{(6)}$ of the Symanzik Lagrangian. For our next-to-leading 
order calculations, we will not require the explicit form of either term. The effective contribution
can be deduced from symmetry considerations alone. 
In the baryon Lagrangian, the mass operator is defined by
\begin{equation}
{\mathcal M}_+ = \frac{1}{2}\left(\xi^\dagger m_Q \xi^\dagger + \xi m_Q \xi\right)
.\end{equation}
Above, the parameter $\D \sim \e \L_\chi$ is the mass splitting between the $\bf{1938}$ and $\bf{1086}$ in the chiral limit.
The parenthesis notation used in Eq.~\eqref{eqn:L} is that of~\cite{Labrenz:1996jy} and is defined 
so that the contractions of flavor indices maintain proper transformations under chiral rotations.

The Lagrangian describing the interactions of the $\cB^{ijk}$ 
and $\cT_\mu^{ijk}$ with the pseudo-Goldstone mesons is
\begin{equation} \label{eqn:Linteract}
  {\cal L} =   
	  2 \a \left(\ol \cB S_\mu \cB A_\mu \right)
	+ 2 \b \left(\ol \cB S_\mu A_\mu \cB \right)
	- 2{\mathcal H}\left(\ol{\cT}_\nu S_\mu A_\mu \cT_\nu\right) 
    	+ \sqrt{\frac{3}{2}}\cC
  		\left[
    			\left(\ol{\cT}_\nu A_\nu \cB\right)+ \left(\ol \cB A_\nu \cT_\nu\right)
  		\right]  
.\end{equation}
The axial-vector and vector meson fields $A_\mu$ and $V_\mu$
are defined by: $ A_\mu=\frac{i}{2}
\left(\xi\partial_\mu\xi^\dagger-\xi^\dagger\partial_\mu\xi\right)$  
and $V_\mu=\frac{1}{2} \left(\xi\partial_\mu\xi^\dagger+\xi^\dagger\partial_\mu\xi\right)$.
The latter appears in  Eq.~\eqref{eqn:L} for the
covariant derivatives of $\cB_{ijk}$ and $\cT_{ijk}$ 
that both have the form
\begin{equation}
  ({\mathcal D}_\mu \cB)_{ijk}
  =
  \partial_\mu \cB_{ijk}
  +(V_\mu)_{il}\cB_{ljk}
  +(-)^{\eta_i(\eta_j+\eta_m)}(V_\mu)_{jm}\cB_{imk}
  +(-)^{(\eta_i+\eta_j)(\eta_k+\eta_n)}(V_\mu)_{kn}\cB_{ijn}
.\end{equation}
The vector $S_\mu$ is the covariant spin operator~\cite{Jenkins:1991jv,Jenkins:1991es}.
The interaction Lagrangian in Eq.~\eqref{eqn:Linteract} also receives lattice spacing corrections. 
In calculating baryon observables, however, these lead to effects that are $\cO(\e^2)$  higher
than the next-to-leading order results.

The parameters that appear in the mixed-action \PQCPT\ Lagrangian can be related to those
in \CPT\ by matching. One realizes that QCD is contained in the fourth-root of the sea-sector of the 
theory. Thus we can relate the parameters of \CPT\ by matching onto terms in the above Lagrangian
restricted to one taste for each flavor of staggered sea quark. For definiteness, we restrict the indices
to $4,8,12$ corresponding to $j_1$, $l_1$, and $r_1$. 
This allows the usual identifications: $\a = \frac{2}{3} D + 2 F$, $\b = - \frac{5}{3} D + F$,
and the remaining parameters: $\a_M$, $\b_M$, $\sigma_M$, $\gamma_M$, $\ol \sigma_M$, $\mathcal{C}$, 
and $\mathcal{H}$, all have the same numerical values as in \CPT.

\section{Baryon Magnetic Moments} \label{s:magmom}

In this Section, we calculate the octet baryon magnetic moments in mixed action \PQCPT. 
We choose to present the calculation of  magnetic moments first due to their simplicity. 
Recall that in the continuum limit, the baryon magnetic moments have the 
behavior~\cite{Jenkins:1992pi,Meissner:1997hn,Durand:1997ya}
\begin{equation}
\mu \sim \mu_0 + \a \sqrt{m_q} + \ldots 
\end{equation}
in the chiral expansion.  
Terms denoted by $\ldots$ scale with a higher power of $m_q$, and are first encountered from one-loop graphs at 
next-to-next-to-leading order. At next-to-leading order then, the magnetic moments receive 
$\cO(\e)$ contributions from loops. 

\begin{figure}
\epsfig{file=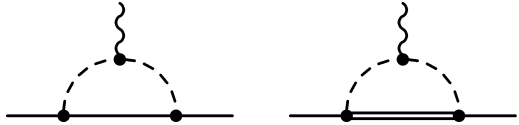}
\caption{Loop diagrams contributing to the octet baryon magnetic moments at $\cO(\e)$. 
The photon is pictured as a wiggly line, mesons are denoted by a dashed line,  
and a thin solid line denotes a $\bm{1938}$ baryon, while the double line denotes a $\bm{1086}$ baryon. 
}
\label{F:amagmom}
\end{figure}

Moving away from the continuum, we are forced to address the corrections from $\cO(a^2)$ operators. 
While there are new $a^2$-dependent terms in the baryon Lagrangian, there are still further terms 
because the vector-current operator also receives $\cO(a^2)$ corrections in the effective theory.  Potential contributions 
from any such terms, however, scale as $\cO(\e^2)$ and are relevant only at next-to-next-to-leading order.
We can safely ignore such contributions.  
This situation is similar to the calculation of 
magnetic moments with mixed actions of Wilson and Ginsparg-Wilson quarks~\cite{Beane:2003xv}, see 
also~\cite{Arndt:2004we}.  Thus the lattice-spacing artifacts at next-to-leading order can only enter
through loop effects.  The one-loop graphs that give a contribution to the octet magnetic moments are
depicted in Figure~\ref{F:amagmom}.

To calculate these diagrams and the tree-level contributions, we must extend
the electric charge matrix $\cQ$, which is not uniquely defined in partially quenched 
theories~\cite{Golterman:2001qj}. By imposing the charge matrix $\cQ$ to be supertraceless
in $SU(15|3)$, no new operators involving the singlet component are introduced. There are various
ways to accomplish this, however, many are not practical to implement time-wise in current lattice computations. 
We require that ghost charges equal their valence counterparts so that operator self-contractions from the 
valence quarks are completely canceled by ghosts~\cite{Tiburzi:2004mv}. Thus our form of the $SU(15|3)$ 
charge matrix is
\begin{equation} \label{eq:Q}
\cQ = \diag (q_u, q_d, q_s, q_j \, \xi_I, q_l \, \xi_I, q_r \, \xi_I, q_u, q_d, q_s)
,\end{equation}
where to maintain supertracelessness $q_j + q_l + q_r = 0$. Taking the fourth-root of the determinant in the continuum, 
and making the sea quarks degenerate with the valence quarks, QCD is recovered only for the specific choice of 
charges: $q_u = q_j = \frac{2}{3}$, and $q_d = q_s = q_l = q_r = - \frac{1}{3}$. Using unphysical 
charges for the valence and sea quarks provides a means to access low-energy constants, and thereby determine the physical 
magnetic moments, for example. Notice the choice, $q_j = q_l = q_r = 0$, is allowed by supertracelessness, 
and would considerably free up computation time by eliminating all operator self-contractions.

In \PQCPT\ the leading contribution to the octet magnetic moments arises from two dimension-$5$ 
operators~\cite{Chen:2001yi}\footnote{%
Here we use $F_{\mu \nu} = \partial_\mu A_\nu - \partial_\nu A_\mu$. 
}
\begin{equation}
\cL = \frac{i e}{2 M_B} 
\left[ 
\mu_\a \left( \ol \cB [S_\mu, S_\nu] \cB \cQ \right)
+ 
\mu_\b \left( \ol \cB [S_\mu, S_\nu] \cQ \cB \right)
\right] F_{\mu \nu}
,\end{equation}
which can be matched onto the \CPT\ Lagrangian upon restricting the baryon
field indices to $4,8,12$, and taking the charges to be their physical values.
In terms of the $SU(3)$ matrix $B$ containing the octet baryons, the corresponding \CPT\ Lagrangian is
\begin{equation}
\cL = \frac{i e}{2 M_B} 
\left[
\mu_D \tr \left( \ol B [S_\mu, S_\nu] \{\cQ, B\} \right)
+
\mu_F \tr \left( \ol B [S_\mu, S_\nu] [\cQ, B] \right)
\right] F_{\mu \nu}
,\end{equation}
and hence we find $\mu_\a = \frac{2}{3} \mu_D + 2 \mu_F$, and $\mu_\b = - \frac{5}{3} \mu_D + \mu_F$ by matching. 
These operators contribute to the magnetic moments at $\cO(\e^0)$. There are additional 
operators that make contributions of $\cO(\e)$ that have identical flavor structure, and differ only by 
the insertion of $\D / \L_\chi$. These operators come with new low-energy constants and 
are allowed because the mass-splitting parameter is a chiral singlet. 
These $\D$-dependent operators also function as counter-terms for the one-loop divergences. 
We shall not keep these operators explicitly, and merely treat the leading low-energy constants as arbitrary linear functions 
of $\D$. The ability to determine coefficients for the $\D$-dependent operators requires the ability 
to vary $\D$, and for this reason we treat the $\D$ dependence only implicitly.

Having spelled out the operators for tree-level contributions to the magnetic moments, we now comment on the computation 
of the loop diagrams. Because flavor-neutral mesons remain charge neutral, even with arbitrary quark charges, 
there are no hairpin contributions at next-to-leading order, see Figure~\ref{F:amagmom}.
Loop graphs containing intermediate state baryons with all valence quarks, or two valence quarks and one ghost quark
are identical to those calculated for $SU(6|3)$ in~\cite{Chen:2001yi} (modulo the different choice of electric charge matrix)
because the ground floor, level $A$, and first floor, level $A$ transform identically in each theory (both are singlets
in the sea sector where the group structure is different).
Finally there are the diagrams with baryons consisting of two valence quarks and one sea quark. The corresponding 
loop mesons are mixed mesons; they have one valence quark and one sea quark. 
The propagation of these modes does not break taste-symmetry. This is because the Lagrangian in the mixed sector 
does not contain explicit taste matrices only the matrix $\tau_3$. Thus the mixed meson propagator
obeys~\cite{Bar:2005tu}
\begin{equation}
\widehat{\Phi_{Q_i Q'} \Phi_{\tilde{Q}' \tilde{Q}_j}} (k^2) 
=  \frac{\d_{Q' \tilde{Q}'} \, \d_{Q \tilde Q} \, \d_{ij}}{k^2 + m_{Q_i Q'}^2}
,\end{equation}
where $Q_i$, $\tilde{Q}_j$ label the staggered quark flavors and tastes, and $Q'$, $\tilde{Q}'$ label the Ginsparg-Wilson 
quark flavors.
The mixed meson mass $m_{Q_i Q'}^2$ is given in Eq.~\eqref{eq:SGW}, and is independent of the quark taste $i$.
Now because there is no taste changing in the loop, and the relevant meson and baryon masses are quark 
taste independent, 
we obtain a factor of four for each staggered quark flavor that can propagate in the loop. This factor is then 
canceled by the fourth-root trick. Explicit calculation verifies this. 
The baryon magnetic moments hence have the same form as in $SU(6|3)$ \PQCPT~\cite{Chen:2001yi}, 
and depend only on one new low-energy constant $C_{\text{mix}}$. This low-energy constant appears in the loop meson masses 
that contain a sea quark and a valence quark. 
 
Combining the tree-level and one-loop graphs, we have
\begin{equation}
\mu = Q \mu_F + \a_D \mu_D + \frac{M_B}{4 \pi f^2} \sum_\phi 
\left\{ 
\b_\phi m_\phi + \b'_\phi \frac{\mathcal{C}^2}{\pi} 
\left[
\cF(m_\phi, \D, \mu) + \frac{5}{3}
\right]
\right\}
,\end{equation}
where we have carried out the spin algebra in $D$-dimensions. 
The non-analytic function appearing in the above expression
is defined by
\begin{equation}
\cF(m,\d,\mu) = \sqrt{\d^2 - m^2} \, \log \frac{\d - \sqrt{\d^2 + m^2 + i \epsilon}}{\d + \sqrt{\d^2 + m^2 + i \epsilon}} 
- \d \log \frac{m^2}{\mu^2} 
.\end{equation}
The coefficients $Q$, and $\a_D$  for the tree-level diagrams are listed in 
Table~\ref{T:treelevel}. Ordinarily $Q$ is the baryon charge; this is no longer the 
case with the quark electric charge matrix $\cQ$ given in Eq.~\eqref{eq:Q}.%
\begin{table}
\caption{\label{T:treelevel}
Tree-level coefficients for the magnetic moments of octet baryons.}
\begin{tabular}{c | c c c}\hline\hline
  & $Q$ & $\a_D$ \\ \hline
$p$ 	  	& $q_u - q_s$ & $q_u + q_s$ \\
$n$ 		& $q_d - q_s$ & $q_d + q_s$ \\
$\S^+$ 		& $q_u - q_d$ & $q_u + q_d$ \\
$\S^0$ 		& $0$ 	      & $q_u + q_d$ \\
$\S^-$ 		& $-q_u + q_d$ & $q_u + q_d$ \\
$\Xi^0$ 	& $-q_d + q_s$ & $q_d + q_s$ \\
$\Xi^-$ 	& $-q_u + q_s$ & $q_u + q_s$ \\
$\Lambda$ 	& $0$         & $\frac{1}{3} (q_u + q_d + 4 q_s)$ \\
$\Lambda \S$ 	& $0$ & $\frac{1}{\sqrt{3}} ( q_u - q_d) $ \\
\end{tabular}
\end{table}
The computed values for the $\b_\phi$, and $\b_\phi'$ coefficients are listed for the octet baryons in 
Tables~\ref{T:p}---\ref{T:Lambda}.
The corresponding values of these coefficients for the $\L\S^0$ transition moment are given in
Table~\ref{T:LambdaSigma0}. 
In each table we have listed the values corresponding to  
loop mesons that has mass $m_\phi$. In these diagrams
there are valence-valence mesons, with masses given in Eq.~\eqref{eq:GWGW}, 
and valence-sea mesons, with masses given in Eq.~\eqref{eq:SGW}.  
If a particular meson is not listed then
the values for $\b_\phi$, and $\b_\phi'$ are zero.

The charges $q_j$ and $q_l$ do not appear explicitly in these tables because, in the isospin
limit, they always come in the combination $q_j + q_l$ and this is identical to $- q_r$. 
Again we remark that the supertracelessness of $\cQ$ is maintained
by the computation-time simplifying choice $q_j = q_l = q_r = 0$. For this choice of charges, one 
is not computing the physical magnetic moments; however, the unphysical moments determined are sensitive to the
physical low-energy constants, hence predictions can be made. 
Additionally one has complete freedom to adjust the charges $q_u$, $q_d$, and $q_s$
as these do not contribute to the supertrace of $\cQ$. One can, for example, choose these charges to 
isolate the low-energy constants at tree-level, see Table~\ref{T:treelevel}, or to simplify the chiral extrapolation
by eliminating particular loop mesons.  
\begin{table*}
\caption{\label{T:p}The coefficients $\b_X$, and $\b_X'$ for the proton.}
\begin{ruledtabular}
\begin{tabular}{c | c c c}
  $\phi$ & $\b_\phi$ & $\b_\phi'$ \\ \hline
  $\pi$ & $-\frac{4}{3} D^2 (q_u - q_d)$ & $ - \frac{1}{6} (q_u - q_d) $ \\
  $ju$ & $-\frac{4}{3} (D^2 + 3 F^2) q_u - 2 ( D - F)^2 q_d - \left(\frac{5}{3} D^2 - 2 D F + 3 F^2 \right) q_r$ 
       & $\frac{1}{18} ( 2 q_u + 4 q_d + 3 q_r )$ \\
  $ru$ & $-\frac{2}{3} (D^2 + 3 F^2) q_u - ( D - F)^2 q_d + \left(\frac{5}{3} D^2 - 2 D F + 3 F^2 \right) q_r$ 
       & $\frac{1}{18} ( q_u + 2 q_d - 3 q_r ) $ \\
\end{tabular}
\end{ruledtabular}
\end{table*}
\begin{table*}
\caption{\label{T:n}The coefficients $\b_\phi$, and $\b_\phi'$ for the neutron.}
\begin{ruledtabular}
\begin{tabular}{c | c c c}
  $\phi$ & $\b_\phi$ & $\b_\phi'$ \\ \hline
  $\pi$ & $\frac{4}{3} D^2 ( q_u - q_d)$ & $\frac{1}{6} ( q_u  - q_d) $ \\
  $ju$ & $-2 (D-F)^2 q_u - \frac{4}{3} ( D^2 + 3 F^2) q_d - \left(\frac{5}{3} D^2 - 2 D F + 3 F^2 \right) q_r$ 
       & $\frac{1}{18} ( 4 q_u + 2 q_d + 3 q_r ) $ \\
  $ru$ & $- ( D - F)^2 q_u  - \frac{2}{3} (D^2 + 3 F^2) q_d + \left(\frac{5}{3} D^2 - 2 D F + 3 F^2 \right) q_r$ 
       & $\frac{1}{18} ( 2 q_u +   q_d - 3 q_r ) $ \\
\end{tabular}
\end{ruledtabular}
\end{table*}
\begin{table*}
\caption{\label{T:Sigmaplus}The coefficients $\b_\phi$, and $\b_\phi'$ for the $\S^+$.}
\begin{ruledtabular}
\begin{tabular}{c | c c c}
  $\phi$ & $\b_\phi$ & $\b_\phi'$\\ \hline
  $K$ & $-\frac{4}{3} D^2 (q_u - q_s)$ & $- \frac{1}{6} (q_u - q_s)$ \\
  $ju$ & $-\frac{2}{3} (D^2 + 3F^2) ( 2 q_u + q_r) $ & $\frac{1}{18} ( 2 q_u  + q_r )$ \\
  $ru$ & $- \frac{2}{3} (D^2 + 3 F^2) (q_u - q_r)$ & $\frac{1}{18} (q_u - q_r) $ \\
  $js$ & $- (D-F)^2 ( 2 q_s + q_r)$ & $ \frac{1}{9} (2 q_s + q_r ) $ \\
  $rs$ & $- (D-F)^2 (q_s - q_r)$ & $ \frac{1}{9} (q_s - q_r ) $ \\
\end{tabular}
\end{ruledtabular}
\end{table*}
\begin{table*}
\caption{\label{T:Sigma0}The coefficients $\b_\phi$, and $\b_\phi'$ for the $\S^0$.}
\begin{ruledtabular}
\begin{tabular}{c | c c c}
  $\phi$ & $\b_\phi$ & $\b_\phi'$ \\ \hline
  $K$ & $-\frac{2}{3} D^2 (q_u + q_d - 2 q_s)$ & $ - \frac{1}{12} (q_u + q_d - 2 q_s )$ \\
  $ju$ & $-\frac{2}{3} ( D^2 + 3 F^2) ( q_u + q_d + q_r)$ & $ \frac{1}{18} (q_u + q_d + q_r) $ \\
  $ru$ & $- \frac{1}{3} (D^2 + 3 F^2) ( q_u + q_d - 2 q_r)$ & $\frac{1}{36} (q_u + q_d - 2 q_r) $ \\
  $js$ & $- (D-F)^2 ( 2 q_s + q_r)$ & $\frac{1}{9} (2 q_s + q_r) $ \\
  $rs$ & $- (D-F)^2 (q_s - q_r) $ & $\frac{1}{9} (q_s - q_r) $ \\
\end{tabular}
\end{ruledtabular}
\end{table*}
\begin{table*}
\caption{\label{T:Sigmaminus}The coefficients $\b_\phi$, and $\b_\phi'$ for the $\S^-$.}
\begin{ruledtabular}
\begin{tabular}{c | c c c}
  $\phi$ & $\b_\phi$ & $\b_\phi'$ \\ \hline
  $K$ & $-\frac{4}{3} D^2 (q_d - q_s)$ & $- \frac{1}{6} (q_d - q_s) $ \\
  $ju$ & $-\frac{2}{3} (D^2 + 3 F^2) (2 q_d + q_r)$ & $\frac{1}{18} (2 q_d + q_r) $ \\
  $ru$ & $-\frac{2}{3} (D^2 + 3 F^2) (q_d - q_r) $ & $\frac{1}{18} (q_d - q_r)$ \\
  $js$ & $- (D-F)^2 ( 2 q_s + q_r)$ & $\frac{1}{9} (2 q_s + q_r) $ \\
  $rs$ & $- (D-F)^2 ( q_s - q_r) $ & $\frac{1}{9} (q_s - q_r) $ \\
\end{tabular}
\end{ruledtabular}
\end{table*}
\begin{table*}
\caption{\label{T:Xi0}The coefficients $\b_\phi$, and $\b_\phi'$ for the $\Xi^0$.}
\begin{ruledtabular}
\begin{tabular}{c | c c c}
  $\phi$ & $\b_\phi$ & $\b_\phi'$ \\ \hline
  $K$ & $\frac{4}{3} D^2 (q_u - q_s)$ & $\frac{1}{6} (q_u - q_s)$ \\
  $ju$ & $- (D-F)^2 ( 2 q_u + q_r)$ & $ \frac{1}{9} (2 q_u + q_r)$ \\
  $ru$ & $- (D-F)^2 ( q_u - q_r)$ & $\frac{1}{9} (q_u - q_r)$ \\
  $js$ & $- \frac{2}{3} ( D^2 + 3 F^2) ( 2 q_s + q_r) $ & $\frac{1}{18} (2 q_s + q_r)$ \\
  $rs$ & $- \frac{2}{3} ( D^2 + 3 F^2) (q_s - q_r) $ & $\frac{1}{18} (q_s - q_r)$ \\
\end{tabular}
\end{ruledtabular}
\end{table*}
\begin{table*}
\caption{\label{T:Ximinus}The coefficients $\b_\phi$, and $\b_\phi'$ for the $\Xi^-$.}
\begin{ruledtabular}
\begin{tabular}{c | c c c}
  $\phi$ & $\b_\phi$ & $\b_\phi'$ \\ \hline
  $K$ & $\frac{4}{3} D^2 ( q_d - q_s)$ & $ \frac{1}{6} ( q_d - q_s)$ \\
  $ju$ & $- (D-F)^2 ( 2 q_d + q_r)$ & $\frac{1}{9} ( 2 q_d + q_r)$ \\
  $ru$ & $- (D-F)^2 ( q_d - q_r)$ & $\frac{1}{9} (q_d - q_r )$ \\
  $js$ & $- \frac{2}{3} (D^2 + 3 F^2) ( 2 q_s + q_ r) $ & $\frac{1}{18} (2 q_s + q_r )$ \\
  $rs$ & $- \frac{2}{3} ( D^2 + 3 F^2) ( q_s - q_r) $ & $\frac{1}{18} (q_s - q_r)$ \\
\end{tabular}
\end{ruledtabular}
\end{table*}
\begin{table*}
\caption{\label{T:Lambda}The coefficients $\b_\phi$, and $\b_\phi'$ for the $\L$.}
\begin{ruledtabular}
\begin{tabular}{c | c c c}
  $\phi$ & $\b_\phi$ & $\b_\phi'$ \\ \hline
  $K$  & $\frac{2}{3} D^2 (q_u + q_d - 2 q_s)$ & $\frac{1}{12} (q_u + q_d - 2 q_s )$ \\
  $ju$ & $-\frac{2}{9} (7D^2 - 12 D F + 9 F^2)(q_u + q_d + q_r)$ & $\frac{1}{6} (q_u + q_d + q_r )$ \\
  $ru$ & $-\frac{1}{9} (7D^2 - 12 D F + 9 F^2)(q_u + q_d - 2 q_r)$ & $\frac{1}{12} (q_u + q_d - 2 q_r) $ \\
  $js$ & $-\frac{1}{9} (D^2 + 6 D F + 9 F^2)(2 q_s + q_r)$ & $0$ \\
  $rs$ & $-\frac{1}{9} (D^2 + 6 D F + 9 F^2)(q_s - q_r)$ & $0$ \\
\end{tabular}
\end{ruledtabular}
\end{table*}
\begin{table*}
\caption{\label{T:LambdaSigma0}The coefficients $\b_\phi$, and $\b_\phi'$ for the $\L\S^0$ transition.}
\begin{ruledtabular}
\begin{tabular}{c | c c c}
  $\phi$ & $\b_\phi$ & $\b_\phi'$ \\ \hline
  $\pi$ & $- \frac{4}{3 \sqrt{3}} D^2 (q_u - q_d)$ & $- \frac{1}{6 \sqrt{3}} ( q_u - q_d) $ \\ 
  $K$ & $- \frac{2}{3 \sqrt{3}} D^2 (q_u - q_d)$ & $- \frac{1}{12 \sqrt{3}} (q_u - q_d ) $ \\
  $ju$ & $\frac{4}{3\sqrt{3}} (D^2 - 3 DF) (q_u - q_d)$ & $- \frac{1}{6 \sqrt{3}} (q_u - q_d) $ \\
  $ru$ & $\frac{2}{3 \sqrt{3}}(D^2 - 3 DF) (q_u - q_d)$ & $ - \frac{1}{12 \sqrt{3}}(q_u - q_d) $ \\
\end{tabular}
\end{ruledtabular}
\end{table*}
The form of the continuum extrapolation of baryon magnetic moments 
is thus highly constrained, especially since the only new low-energy constant
at this order, $C_{\text{mix}}$, could be determined independently 
from the masses of mixed mesons~\cite{Bar:2005tu}.

\section{Baryon Masses} \label{s:mass}

\begin{figure}
\epsfig{file=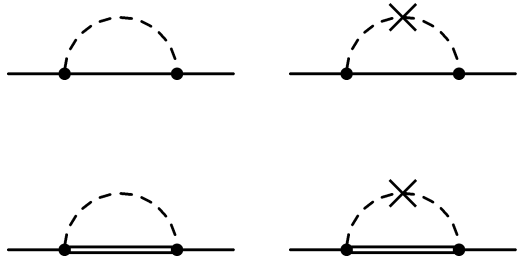}
\caption{Loop diagrams contributing to the octet baryon masses at $\cO(\e^3)$. 
Mesons are denoted by a dashed line, flavor neutrals (hairpins) by a crossed dashed line, 
and a thin solid line denotes a $\bm{1938}$ baryon, while the double line denotes a $\bm{1086}$ baryon. 
}
\label{F:amass}
\end{figure}

In this Section we determine the octet baryon masses in mixed action \PQCPT.
Let us first recall the behavior of the baryon mass in the continuum, and near the chiral 
limit~\cite{Gasser:1980sb,Jenkins:1991ts,Jenkins:1991bs,Bernard:1993nj}
\begin{equation}
M_B \sim M_0 + \a \, m_q + \b \, m_q^{3/2} + \ldots
,\end{equation}
where we have retained the leading non-analytic piece and the $\ldots$ denotes terms with
higher powers of the quark mass. The non-analytic term above
stems from the one-loop diagrams shown in Figure~\ref{F:amass}.
A mass calculation at next-to-leading order is $\cO(\e^3)$ in our power counting.
To perform a complete  $\cO(\e^3)$ calculation, we
must retain the $a^2$-dependent piece of the loop meson masses and 
evaluate the potential $\mathcal{V}_\cB$ in Eq.~\eqref{eqn:L} at tree level. Remember that 
there are no $a^3$ terms in the Lagrangian. 

Without writing down the explicit form of the
baryon potential, we can deduce the net contribution from $\mathcal{V}_\cB$ to the octet masses at tree level. 
Because the external states all involve Ginsparg-Wilson quarks, the action of the potential
is necessarily a taste-singlet at tree level. Furthermore since insertion of the mass matrix is
at $\cO(\e^4)$, there are no valence flavor matrices around. This implies that the evaluation of 
the baryon potential is identical for all octet states: in essence all Ginsparg-Wilson quarks
are identical in the potential at this order. While there are terms in the potential that
violate $SO(4)$ rotational invariance~\cite{Tiburzi:2005vy}, the mass only picks up an 
indiscernible additive shift from such terms [it is the dispersion relation that is sensitive
to $SO(4)$ breaking].  Therefore the effect of the potential at $\cO(a^2)$ can be described by 
just one operator
\begin{equation}
\mathcal{V}_{\cB} \overset{\text{eff}}{=} - C_0 \left( \ol \cB \cB \right) 
.\end{equation}
This term and the familiar tree-level terms that are linear in the quark masses are straightforwardly evaluated.
We must keep in mind that $\a_M$, $\b_M$, and $\sigma_M$ are to be treated as arbitrary linear
functions of $\D / \L_\chi$. The linear terms, for example, function in part as counter-terms for the one-loop
diagrams.

Now we turn to the loop diagrams in Figure~\ref{F:amass}. 
As with the computation of the magnetic moments, the diagrams involving baryons with 
three valence quarks, or two valence quarks and one ghost quark are identical to those
in $SU(6|3)$ \PQCPT~\cite{Chen:2001yi}, because these particular floors and levels 
of each multiplet are identical. 
Furthermore, diagrams with one sea quark flowing in the loop are evaluated in 
precisely the same way as before. The mixed-meson propagators are diagonal in taste and flavor,
while the meson and baryon masses are taste-independent. Consequently the loops come with a four-fold 
degeneracy when one sums over the sea-quark tastes. This degeneracy is canceled by the factor of 
$1/4$ to implement the fourth-root trick. The only new contribution as compared to the magnetic moments 
is that of flavor-neutral modes. But at this order, the flavor-neutral mesons which propagate in loops
are those consisting of two Ginsparg-Wilson valence quarks. These are necessarily taste-singlets;
their propagators appear in Eq.~\eqref{eq:singlet}, and involve taste-singlet sea-sea meson masses, which are 
$a^2$-dependent.

Assembling the tree-level results with the loop-diagrams, we find
\begin{eqnarray}
M_B &=& M_0 - 2 m_u C_u - 2 m_s C_s - 2 \sigma_M ( 2 m_j + m_r ) - a^2 C_0 \notag \\
&& 
- \frac{1}{8 \pi f^2} 
\left[ \sum_\phi A_\phi m_\phi^3 + \sum_{\phi \phi'} A_{\phi \phi'} \cM^3(m_\phi, m_{\phi'}) \right]
\notag \\
&&
- \frac{\mathcal{C}^2}{8 \pi^2 f^2}
\left[
\sum_\phi B_\phi F(m_\phi, \D, \mu) + 
\sum_{\phi, \phi'} B_{\phi \phi'} F(m_\phi, m_{\phi'}, \D, \mu)
\right]
.\end{eqnarray}
The non-analytic functions appearing in the expression for the octet baryon masses are defined by
\begin{equation}
F(m,\d,\mu) = (m^2 - \d^2) 
\left( 
\sqrt{\d^2 - m^2} \, \log \frac{\d - \sqrt{\d^2 + m^2 + i \epsilon}}{\d + \sqrt{\d^2 + m^2 + i \epsilon}} 
- \d \log \frac{m^2}{\mu^2} 
\right)
- \frac{1}{2} \d m^2 \log \frac{m^2}{\mu^2}
,\end{equation}
and
\begin{eqnarray}
\cM^3( m_\phi, m_{\phi'}) &=& \cH_{\phi \phi'}(m_\phi^3, m_{\phi'}^3, m_X^3), \\
F(m_\phi, m_{\phi'}, \d, \mu) &=& \cH_{\phi \phi'}[F(m_\phi, \d, \mu), F(m_{\phi'}, \d, \mu), F(m_X, \d, \mu)] 
,\end{eqnarray}
for the flavor-neutral contributions.

In Table~\ref{T:treemass}, we list the coefficients $C_u$ and $C_s$
for the octet baryons. 
\begin{table}
\caption{\label{T:treemass}
Tree-level coefficients for the octet baryon masses.}
\begin{tabular}{c | c c c}\hline\hline
  & $C_u$ & $C_d$ \\ \hline
$N$ & $\a_M + \b_M$ & $0$ \\
$\Sigma$ & $\frac{1}{6}(5 \a_M + 2 \b_M)$ & $\frac{1}{6} (\a_M + 4 \b_M)$ \\
$\Lambda$ & $\frac{1}{2} ( \a_M + 2 \b_M)$ & $\frac{1}{2} \a_M$ \\
$\Xi$ & $\frac{1}{6}(\a_M + 4 \b_M) $ & $\frac{1}{6} ( 5 \a_M + 2 \b_M)$ \\
\end{tabular}
\end{table}
In Tables~\ref{t:PQQCD-A} and \ref{t:PQQCD-B}, we list the loop coefficients $A_\phi$, $A_{\phi, \phi'}$, 
$B_{\phi}$ and $B_{\phi\phi'}$. These coefficients are grouped according to loop mesons with mass
$m_\phi$, and for flavor-neutral contributions are grouped according to pairs of quark-basis flavor-neutral mesons. 
Notice the sum on $\phi\phi'$ runs over $\eta_u \eta_u$, $\eta_u \eta_s$, and $\eta_s \eta_s$ to avoid double counting.

The lattice spacing dependence of the 
octet masses too is highly constrained: there are only three free parameters: $C_0$, which is the same for all octet
baryons; 
$C_{\text{mix}}$, which alternately can be determined from mixed meson masses; and the combination of parameters $C_3 + C_4$, which 
is already constrained from staggered meson lattice data~\cite{Aubin:2004fs}.

\begin{table}
\caption{The coefficients $A_\phi$ and $A_{\phi\phi'}$ in \PQCPT. Coefficients are
listed for the baryon octet, and for $A_\phi$ are grouped into contributions from loop mesons
with mass $m_\phi$, while for $A_{\phi\phi'}$ are grouped into contributions from pairs of quark-basis 
$\eta_q$ mesons.}
\begin{tabular}{l | c c c c }
 & \multicolumn{4}{c}{$A_\phi \phantom{ap}$} \\
    & $\quad \pi \quad$ & $\quad K \quad $ & $\quad \eta_s \quad $ & \\
\hline
$N $       &  $-\frac{4}{3}(D^2 - 3 D F)$ & $0$  & $0$  & \\

$\Sigma$   &  $-\frac{2}{3} (D^2 - 3 F^2)$ & $-\frac{2}{3} (D^2 - 6 D F + 3 F^2)$  & $0$  & \\

$\Lambda$  &  $-\frac{2}{9} (D^2 - 12 D F + 9 F^2)$ & $-\frac{2}{9}(5 D^2 - 6 D F - 9 F^2)$  & $0$  & \\

$\Xi$      &  $0$ & $-\frac{2}{3} (D^2 - 6 D F + 3 F^2)$  & $-\frac{2}{3}(D^2 - 3F^2)$  & \\
         
\multicolumn{5}{c}{} 
\\
 	& $\quad ju \quad$ & $\quad ru \quad$ & $\quad js \quad$ & $\quad rs \quad$ \\
\hline
$N$     &  $\frac{2}{3}(5 D^2 - 6 D F + 9 F^2)$ & $\frac{1}{3}(5 D^2 - 6 D F + 9 F^2)$  
           &  $0$ & $0$ \\
$\Sigma$            &  $\frac{4}{3} (D^2 + 3 F^2)$ & $\frac{2}{3} (D^2 + 3 F^2)$  
           &  $2 (D-F)^2$ & $(D-F)^2$ \\
$\Lambda$ 
           &  $\frac{4}{9} (7 D^2 - 12 D F + 9 F^2)$ & $\frac{2}{9} (7 D^2 - 12 D F + 9 F^2)$  
           &  $\frac{2}{9} ( D + 3 F)^2$ & $\frac{1}{9} ( D + 3 F)^2$ \\
$\Xi$   &  $2 ( D - F)^2 $ & $(D -F)^2$  
           &  $\frac{4}{3} (D^2 + 3 F^2)$ & $\frac{2}{3} (D^2 + 3 F^2)$ \\           
\multicolumn{5}{c}{} 
\\
& \multicolumn{4}{c}{$A_{\phi\phi'}$ \phantom{sp}} 
\\
    	& $\quad \eta_u \eta_u \quad$ & $\quad \eta_{u} \eta_s \quad $& $\quad \eta_s \eta_s \quad$ &\\
\hline
$N$            &  $(D-3F)^2$ & $0$ & $0$ \\
$\Sigma$            &  $4 F^2$ & $- 4 (D F - F^2)$ & $(D - F)^2$ \\
$\Lambda$ &  $\frac{4}{9} (2 D - 3 F)^2$ & $- \frac{4}{9} (2 D^2 + 3 D F - 9 F^2)$ & $\frac{1}{9}(D + 3 F)^2$ \\
$\Xi$            &  $(D - F)^2$ & $- 4 (D F - F^2)$ & $4 F^2$ \\
\end{tabular}
\label{t:PQQCD-A}
\end{table}

\begin{table}
\caption{The coefficients $B_\phi$ and $B_{\phi\phi'}$ in \PQCPT. Coefficients are
listed for the octet baryons, and for $B_\phi$ are grouped into contributions from loop mesons
with mass $m_\phi$, while for $B_{\phi\phi'}$ are grouped into contributions from pairs of quark-basis 
$\eta_q$ mesons.}
\begin{tabular}{l | c c c c c c c | c c c }
 & \multicolumn{7}{c|}{$B_\phi \phantom{ap}$} & \multicolumn{3}{c}{$B_{\phi\phi'}$ \phantom{sp}} \\
    & $\quad \pi \quad$ & $\quad K \quad $ & $\quad \eta_s \quad $ 
    & $ \quad ju \quad$ & $ \quad ru \quad$ 
    & $\quad js \quad$  & $\quad rs \quad$ 
    & $\quad \eta_u \eta_u \quad $ & $\quad \eta_u \eta_s\quad $   & $\quad \eta_s \eta_s\quad$ \\
\hline
$N$        &  $\frac{2}{3}$ & $0$  & $0$  
           &  $\frac{2}{3}$ & $\frac{1}{3}$  
           &  $0$ & $0$
           &  $0$ & $0$ & $0$ \\

$\Sigma$   &  $\frac{1}{9}$ & $\frac{5}{9}$  & $0$  
           &  $\frac{2}{9}$ & $\frac{1}{9}$  
           &  $\frac{4}{9}$ & $\frac{2}{9}$
           &  $\frac{2}{9}$ & $- \frac{4}{9}$ & $\frac{2}{9}$ \\

$\Lambda$  &  $\frac{1}{3}$ & $\frac{1}{3}$  & $0$  
           &  $\frac{2}{3}$ & $\frac{1}{3}$  
           &  $0$ & $0$
           &  $0$ & $0$ & $0$ \\

$\Xi$      &  $0$ & $\frac{5}{9}$  & $\frac{1}{9}$  
           &  $\frac{4}{9}$ & $\frac{2}{9}$  
           &  $\frac{2}{9}$ & $\frac{1}{9}$
           &  $\frac{2}{9}$ & $-\frac{4}{9}$ & $\frac{2}{9}$ \\

\end{tabular}
\label{t:PQQCD-B}
\end{table}

\section{\label{s:summy}Summary}

Above we have included baryons into mixed-action partially-quenched chiral perturbation theory
for Ginsparg-Wilson valence quarks and staggered sea quarks. Working at next-to-leading order, 
we determined the lattice spacing artifacts for the octet baryon magnetic moments and masses.
The recipe for adding finite volume corrections is discussed in the Appendix. 

To $\cO(\e^3)$, baryon masses depend on the lattice spacing via three parameters. The first is $C_0$, which is a 
representative coefficient of local $a^2$ operators in the baryon Lagrangian. The contribution from $C_0$ is the same
for all members of the baryon octet. The second parameter is $C_{\text{mix}}$, which affects the masses of mesons 
made from a Ginsparg-Wilson quark and a staggered quark. The third parameter is a combination of low-energy constants,
$C_3 + C_4$, which governs the mass of taste-singlet staggered mesons.   
To $\cO(\e)$, baryon magnetic moments only depend on the lattice spacing through loop-meson masses.
Furthermore there is only one free parameter involved at next-to-leading order, $C_{\text{mix}}$.

A simplifying feature of the mixed action baryon theory is that at next-to-leading 
order, there is no taste-symmetry violation.  Beyond next-to-leading order, however, 
this is no longer the case. Taste-symmetry breaking interactions start at $\cO(\e^4)$ in our power counting. 
They arise from two sources. Contributions that scale as $\cO(m_q^2 \log m_q)$ can lead to taste-symmetry violation. 
Expanding out the sigma terms in the baryon Lagrangian~\eqref{eqn:L} leads to contributions to the baryon masses that 
scale as $m_q^2 \log m_q$, and involve completely disconnected meson loops. In \PQCPT\ these loops 
involve only mesons formed from two sea-quarks~\cite{Walker-Loud:2004hf,Tiburzi:2004rh,Tiburzi:2005na}. 
For staggered sea quarks this becomes a sum over the various meson tastes (vector, axial-vector,...)
and no longer is there a cancellation of factors of $1/4$ inserted from the fourth-root trick. 
Explicit taste-symmetry violation occurs from operators that scale as $\cO(a^2 m_q \log m_q)$.
These arise from the loop-diagrams generated by terms in $\mathcal{V}_\cB$ (for which one would need to decompose $C_0$
into various contributing terms). Terms in $\mathcal{V}_\cB$ containing taste-spurions will generate taste-symmetry 
violation at the one-loop level. 
These are the only possible sources for taste-symmetry violation at next-to-next-to-leading order. 
The remainder of contributions to the baryon masses and other observables arise from 
valence-valence or valence-sea mesons in the absence of explicit taste matrices. 
These do not violate taste-symmetry as we demonstrated above.

Nonetheless the issue of addressing lattice spacing corrections to baryon observables calculated in 
mixed-action lattice QCD is very tractable. The chiral symmetry properties of Ginsparg-Wilson valence
quarks effectively suppress the taste-symmetry violation from the staggered sea. While taste-symmetry violation
does occur at next-to-next-to-leading order, expressions at leading-order are taste-symmetric and 
involve only a few new parameters.
Mixed-action simulations in the baryon sector are a timely way of getting physical observables from lattice QCD.

\begin{acknowledgments}
We thank Andr\'e Walker-Loud for critical discussions during the early stages of this work,
and Tom Mehen for reading a draft of the manuscript.
This work is supported in part by the U.S.\ Department of Energy under Grant No.\ DE-FG02-96ER40945. 
\end{acknowledgments}

\appendix

\section*{Finite Volume Corrections}

In this Appendix, we assemble the relevant formulae for finite volume corrections. 
This is a trivial extension of Ref.~\cite{Beane:2004tw}, and is included for completeness.  
Let $L$ denote the size of the cubic box in one spatial direction. 
We assume the lattice simulations are carried out in a region of parameter space where
chiral physics lives inside the box, i.e. for $f L \gg 1$, and further that we are in the $p$-regime
of chiral perturbation theory, $m_\pi L \gtrsim 1$. In this case, the Poisson formula can be used 
to cast mode sums from loop diagrams into the infinite volume results plus finite volume modifications. We list these
modifications for the octet baryon masses and magnetic moments. For finite volume corrections in mixed action simulations, we imagine that the Ginsparg-Wilson valence quarks
will be the lightest. Thus modulo possible cancellations from additive $a^2$ mass shifts for the valence-sea and 
sea-sea mesons, the valence-valence pions should dominate the finite volume corrections. 

The finite volume corrections to the magnetic moments have the form~\cite{Beane:2004tw}
\begin{equation}
\d_L \mu = - \frac{M_B}{6 \pi^2 f^2} 
\sum_\phi \left[ 
\b_\phi \cY(m_\phi, 0)
+ 
\b'_\phi \mathcal{C}^2 \cY(m_\phi, \D)
\right]
,\end{equation}
where
\begin{equation}
\cY(m, \D) = \int_0^\infty d\lambda \sum_{\bm{n}- \bm{0}} 
\Big[ 
3 K_0 (\b_\D | \bm{n} | L) - \b_\D |\bm{n}| L \; K_1 ( \b_\D |\bm{n}| L )
\Big]
,\end{equation}
with $\b_\D^2 = m^2 + 2 \D \lambda + \lambda^2$ and the $K_n(x)$ are modified Bessel functions. 
The coefficients $\b_\phi$ and $\b'_\phi$ are listed for the octet baryons in Tables~\ref{T:p}--\ref{T:Lambda}. 
The coefficients for the $\Lambda \Sigma^0$transition moment appear in Table~\ref{T:LambdaSigma0}.

The finite volume corrections to the masses have the form~\cite{Beane:2004tw}
\begin{eqnarray}
\d_L M_B &=& -\frac{1}{4 \pi f^2} \left[ \sum_\phi A_\phi \cK (m_\phi, 0) + 
\sum_{\phi \phi'} A_{\phi \phi'} \cK(m_\phi, m_{\phi'}, 0) \right]
\notag \\
&& - 
\frac{\mathcal{C}^2}{4 \pi^2 f^2} \left[ 
\sum_\phi B_\phi \cK(m_\phi, \D) + \sum_{\phi \phi'} B_{\phi \phi'} \cK(m_\phi, m_{\phi'}, \D)
\right],
\end{eqnarray}
where
\begin{equation}
K(m,\D) = \int_0^\infty d\lambda \b_\D^2 \sum_{\bm{n} - \bm{0}} 
\left[ \frac{K_1(\b_\D |\bm{n}| L )}{\b_\D |\bm{n}| L } - K_0 ( \b_\D |\bm{n}| L )\right]
,\end{equation}
and
\begin{equation}
\cK(m_\phi, m_{\phi'},\D) = \cH_{\phi \phi'} \left[ \cK(m_\phi, \D), \cK(m_{\phi'},\D), \cK(m_X, \D) \right]
.\end{equation} 
The coefficients $A_\phi$ and $A_{\phi,\phi'}$ are listed for the octet baryons 
in Table~\ref{t:PQQCD-A}, while $B_\phi$ and $B_{\phi \phi'}$ appear in Table~\ref{t:PQQCD-B}.

\bibliography{hb}

\end{document}